\RequirePackage{fixltx2e}
\documentclass[a4paper,fleqn,reqno,english,journal=cmatex,manuscript=article]{revtex4-2}
\usepackage[T1]{fontenc}
\usepackage[latin9]{inputenc}
\usepackage{babel}
\usepackage{array}
\usepackage{float}
\usepackage{textcomp}
\usepackage{multirow}
\usepackage{amsmath}
\usepackage{amsthm}
\usepackage{amssymb}
\usepackage{stmaryrd}
\usepackage{graphicx}
\usepackage[unicode=true,pdfusetitle,
 bookmarks=true,bookmarksnumbered=false,bookmarksopen=false,
 breaklinks=false,pdfborder={0 0 1},backref=false,colorlinks=false]
 {hyperref}

\makeatletter

\newcommand{\lyxmathsym}[1]{\ifmmode\begingroup\def\b@ld{bold}
  \text{\ifx\math@version\b@ld\bfseries\fi#1}\endgroup\else#1\fi}

\providecommand{\tabularnewline}{\\}

\usepackage{lmodern}
\makeatother

\begin{document}
\title{Exploring 2D/Quasi-2D Ruddlesden\textendash Popper Perovskite A$_{n+1}$Hf$_{n}$S$_{3n+1}$
(A = Ca, Sr, and Ba; $n$ = 1$-$3) for Optoelectronics using Many-Body
Perturbation Theory}
\author{Surajit Adhikari{*}, Aksaj Bharatwaj, and Priya Johari}
\email{sa731@snu.edu.in, priya.johari@snu.edu.in}

\affiliation{Department of Physics, School of Natural Sciences, Shiv Nadar Institution
of Eminence, Greater Noida, Gautam Buddha Nagar, Uttar Pradesh 201314,
India.}
\begin{abstract}
Dimensionality engineering in A$_{n+1}$B$_{n}$X$_{3n+1}$ Ruddlesden$-$Popper
(RP) perovskite phases has emerged as a promising strategy to enhance
optoelectronic properties. These properties are highly material-dependent,
requiring detailed exploration of electronic, optical, excitonic,
transport, and polaronic characteristics. However, the absence of
comprehensive studies continues to impede the rational design of high-performance
materials. In this work, we investigate the excitonic and polaronic
effects in A$_{n+1}$Hf$_{n}$S$_{3n+1}$ (A = Ca, Sr, and Ba; $n$
= 1$-$3) RP phases, examining their relative stability and optoelectronic
properties using several first-principles based methodologies within
the framework of density functional theory and many-body perturbation
theory (GW and BSE). Our study suggests that these compounds are mechanically
stable and feature G$_{0}$W$_{0}$@PBE bandgaps ranging from 1.43
to 2.14 eV, which are smaller than those of their bulk counterparts.
BSE and model-BSE (mBSE) calculations indicate that these RP phases
display notable optical anisotropy, with the exciton binding energy
decreasing as the thickness of the perovskite layer increases. In
addition, intermediate to strong carrier-phonon scattering is observed
in these compounds, confirmed through the Fr\"ohlich mechanism near
room temperature. Using the Feynman polaron model, the polaron parameters
of these RP phases are also computed, and it is found that charge-separated
polaronic states are less stable than bound excitons. Finally, a significant
increase in electron mobilities is observed in RP phases compared
to their bulk counterparts. Overall, the insights gained from this
study will enable the rational design of layered perovskite phases
for applications in solar cells and other optoelectronic devices.
\end{abstract}
\maketitle

\section{Introduction:}

Lead halide perovskites APbX$_{3}$ {[}A = Cs$^{+}$, CH$_{3}$NH$_{3}^{+}$,
CH (NH$_{2}$)$_{2}^{+}$; X = Cl$^{-}$, Br$^{-}$, I$^{-}${]} have
gained immense attention for their applications in optoelectronic
and solar cell, owing to their their exceptional electronic and optical
properties, high carrier mobility, low production cost, and outstanding
power conversion efficiency \citep{chapter2-12,chapter2-53,chapter3-8,chapter3-9,chapter3-10}.
However, the inclusion of organic components in these materials compromises
their stability, making them susceptible to degradation from heat,
light, and moisture, which adversely affects their long-term performance.
Additionally, the toxicity of lead raises serious environmental and
health concerns, further limiting their practical applications \citep{chapter3-12,chapter1-16}.

Seeking alternatives to overcome the limitations of lead halide perovskites,
chalcogenide perovskites (CPs) with sulfur (S) or selenium (Se) anions
have emerged as promising candidates for photovoltaic applications
\citep{chapter3-11,chapter3-16,chapter3-17,chapter3-18}. Several
prototypical distorted CPs, such as ABS$_{3}$ (A = Ca, Sr, Ba; B
= Zr, Hf), have been successfully synthesized and theoretically explored
for photovoltaic applications \citep{chapter1-63,chapter3-19,chapter3-37}.
Among these, Hf-based compounds exhibit higher bandgaps (2.17$-$2.46
eV), lower charge carrier mobility (9.44$-$15.68 cm$^{2}$V$^{-1}$s$^{-1}$),
and reduced power conversion efficiency (10.56$-$13.26\%), which
collectively constrain the photovoltaic performance of Hf-based CPs
\citep{chapter3-19,chapter3-37}. To enhance solar cell performance,
alloying/doping at the Hf and S sites has been explored in these materials
\citep{chapter6-3,chapter6-6,chapter6-7}. Alloying is a widely used
method to modulate the bandgap of semiconductors; however, the resulting
disorder often leads to a reduction in carrier mobility \citep{chapter7-4,chapter7-5}.
Furthermore, doped or alloyed configurations tend to encounter stability
issues \citep{chapter5-13}.

Ruddlesden$-$Popper (RP) phases offer an alternative strategy for
achieving long-range ordered structures, enabling precise tuning of
electronic and optical properties \citep{chapter7-6}. The RP phases,
resembling the perovskite structure, are developing as semiconductors
for optoelectronic applications \citep{chapter7-6,chapter7-7}. The
general formula of RP phases is A$_{n+1}$B$_{n}$X$_{3n+1}$ or AX{[}ABX$_{3}${]}$_{n}$,
where perovskite structure blocks with a unit cell thickness of \textquotedblleft $n$\textquotedblright{}
are separated by a rock salt layer (AX) along the {[}001{]} direction
\citep{chapter7-6}. The alternate perovskite blocks are displaced
in the in-plane direction by half of the unit cell. One of the key
advantages of RP phases is the tunable dimensionality parameter, $n$,
which offers enhanced flexibility in modifying the bandgap, chemical
stability, and charge transport properties, compared to their bulk
conuterparts ABX$_{3}$ ($n=\infty$) \citep{Intro25,chapter7-7,chapter7-6,chapter7-2}.
The bandgap is particularly crucial in determining the practical performance
and efficiency of RP phases in optoelectronic devices, which has led
to extensive research aimed at modulating the bandgap through dimensionality
engineering. 

The RP phases are classified under the broad category of \textquotedblleft 2D
perovskites\textquotedblright{} due to their layered structural arrangement,
where the periodic stacking of perovskite layers creates a bulk structure
(see Figure \ref{Figure:1}). Their material properties can be tuned
through substitution or dimensional reduction. For example, previous
theoretical study indicated that the bandgaps of Ba$_{n+1}$Zr$_{n}$S$_{3n+1}$
($n$ = 1$-$3) RP phases are significantly lower than that of the
parent perovskite, BaZrS$_{3}$ \citep{chapter7-6,Intro25}. Due to
quantum confinement effects, reducing the dimensionality of the material
leads to significant changes in its physical properties and the evolution
of the bandgap \citep{chapter7-6,chapter7-8}.

Inspired by the dimensionality engineering of RP phases for developing
efficient materials for photovoltaic and other optoelectronic applications,
this work aims to investigate the optoelectronic properties along
with excitonic and polaronic effects in A$_{n+1}$Hf$_{n}$S$_{3n+1}$
(A = Ca, Sr, and Ba; $n$ = 1$-$3) RP phases within the framework
of density functional theory (DFT) \citep{chapter2-36,chapter2-37}
and many-body perturbation theory (MBPT) \citep{chapter3-1,chapter3-2}.
Initially, we optimized the crystal structures using the semilocal
PBE exchange-correlation (xc) functional \citep{chapter1-34} and
assessed their stability. To investigate the electronic properties,
band structures are calculated using both the HSE06 functional \citep{chapter1-35}
and the G$_{0}$W$_{0}$@PBE method \citep{chapter1-69,chapter1-70}.
Subsequently, the optical properties are determined using the Bethe$-$Salpeter
equation (BSE) \citep{chapter1-67,chapter1-68} and model-BSE (mBSE)
\citep{chapter1-65} methods. In addition, the BSE/mBSE method and
the Wannier-Mott model \citep{chapter7-3}are employed to compute
the excitonic properties of these compounds, as exciton formation
in optoelectronic materials significantly impacts their charge-separation
characteristics. Polaronic effects are also suggested to play a crucial
role in excitation dynamics and carrier transport. Consequently, electron-phonon
coupling is addressed using the Frohlich model, and polaron mobility
is computed using the Hellwarth polaron model \citep{chapter2-22}.
Overall, this paper demonstrates that dimensionality engineering in
A$_{n+1}$Hf$_{n}$S$_{3n+1}$ (A = Ca, Sr, and Ba; $n$ = 1$-$3)
RP phases offers a promising strategy for developing efficient materials
for optoelectronic devices.

\section{Computational Details:}

The first-principles calculations were carried out using the projected
augmented wave (PAW) \citep{chapter1-33} method as implemented in
the Vienna Ab initio Simulation Package (VASP) \citep{chapter1-31,chapter1-32}.
The generalized gradient approximation (GGA) based Perdew-Burke-Ernzerhof
(PBE) exchange$-$correlation (xc) functional \citep{chapter1-34}
was employed for structural optimizations, effectively incorporating
electron$-$electron interactions. The valence electronic configurations
considered for Ca, Sr, Ba, Hf, and S were 3s$^{2}$3p$^{6}$4s$^{2}$,
4s$^{2}$4p$^{6}$5s$^{2}$, 5s$^{2}$5p$^{6}$6s$^{2}$, 5p$^{6}$6s$^{2}$5d$^{2}$,
and 3s$^{2}$3p$^{4}$, respectively. A $2\times2\times2$ $\Gamma$-centered
$k$-point mesh was utilized, and the plane-wave cutoff energy was
set to 400 eV for optimization calculations. Full structural relaxation
was performed with the electronic self-consistent-field (SCF) iteration
energy convergence criterion of $10^{-6}$ eV and the Hellmann\textminus Feynman
force convergence criterion set to 0.01 eV/$\textrm{\AA}$. The crystal
structures were visualized using the VESTA software \citep{chapter2-3}.
The advanced hybrid HSE06 \citep{chapter1-35} xc functional and MBPT-based
GW (G$_{0}$W$_{0}$@PBE) \citep{chapter1-69,chapter1-70} calculations
were performed to obtain a more precise estimation of the electronic
bandgap. The effective mass was calculated using the Sumo-bandstat
program \citep{chapter2-10}. To explore the optical properties and
excitonic effects, the Bethe$-$Salpeter equation (BSE) \citep{chapter1-67,chapter1-68}
was solved on top of the G$_{0}$W$_{0}$@PBE calculation, explicitly
incorporating the electron-hole interaction. A $\Gamma$-centered
$2\times2\times2$ $k$-grid and a converged NBANDS of 800 were employed
for the GW-BSE calculations. The electron$-$hole kernel for the BSE
calculations was constructed using 8 occupied and 8 unoccupied bands.
However, for A$_{4}$Hf$_{3}$S$_{10}$ (A = Ca, Sr, Ba) compounds,
performing BSE calculations was infeasible due to computational limitations,
even with the fastest supercomputers. To address this challenge, a
less computationally demanding yet reliable model-BSE (mBSE) \citep{chapter1-65}
approach was utilized for these compounds. Following this, the ionic
contribution to the dielectric constant was calculated using the DFPT
method \citep{chapter1-60}.

\section{Results and Discussions:}

In this study, we conducted a comprehensive and systematic investigation
of the Ruddlesden$-$Popper (RP) phases A$_{n+1}$Hf$_{n}$S$_{3n+1}$
(A = Ca, Sr, and Ba; $n$ = 1$-$3) to explore their potential optoelectronic
features. The subsequent sections present an in-depth analysis of
their stability, as well as their structural, electronic, optical,
excitonic, and polaronic properties, providing valuable insights to
guide and inspire future experimental research.

\begin{figure}[H]
\begin{centering}
\includegraphics[width=1\textwidth,height=1\textheight,keepaspectratio]{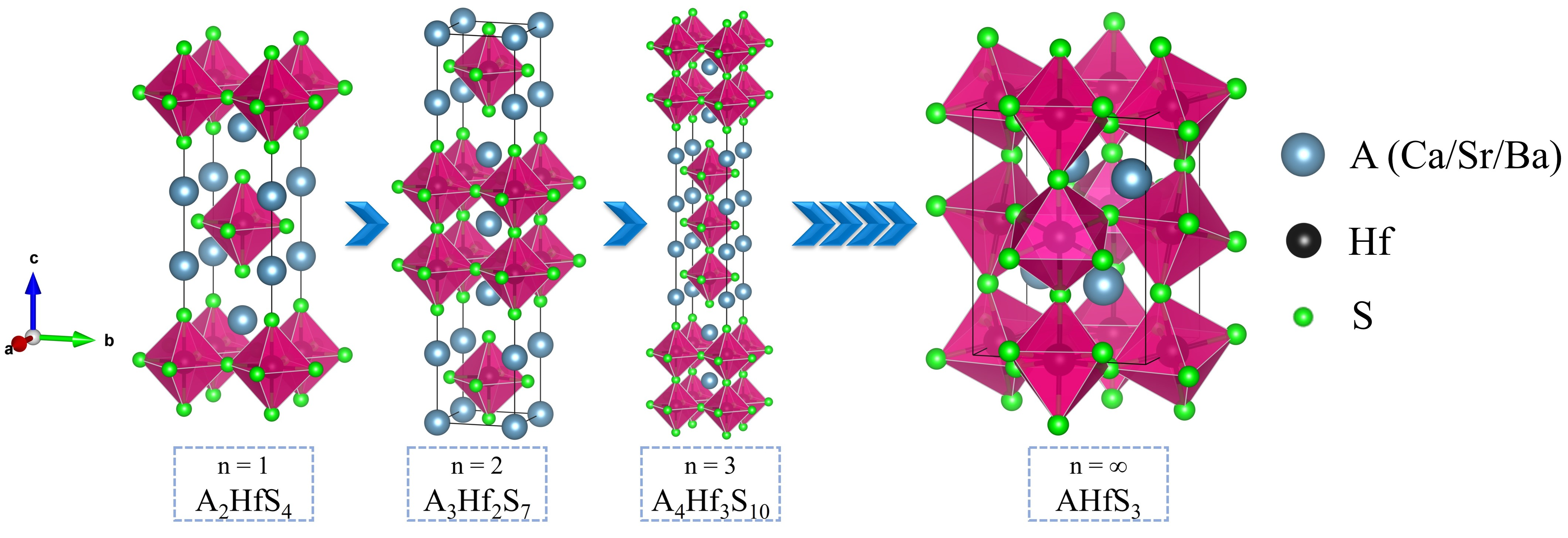}
\par\end{centering}
\caption{\label{Figure:1}Optimized crystal structure of A$_{n+1}$Hf$_{n}$S$_{3n+1}$
(A = Ca, Sr, Ba and $n$ = 1$-$3) RP phases.}
\end{figure}

\subsection{Structural Properties:}

Ruddlesden$\lyxmathsym{\textendash}$Popper phases of chalcogenide
perovskites follow the general formula AX{[}ABX$_{3}${]}$_{n}$,
where ABX$_{3}$ perovskite blocks (P-blocks) are stacked along the
$[001]$ direction. For every $n$ perovskite units, an additional
AX layer is inserted, and the interlayer A cations coordinate with
the terminal chalcogen (X) atoms, forming rock-salt blocks (R-blocks)
\citep{chapter7-1}. Figure \ref{Figure:1} illustrates the crystal
structures of the A$_{n+1}$Hf$_{n}$S$_{3n+1}$ (A = Ca, Sr, Ba and
$n$ = 1$-$3) RP phases. The calculated lattice parameters of these
RP phases are tabulated in Table \ref{Table:1}, confirm that they
adopt a tetragonal structure with the space group $I4/mmm$ (No. 139).
To validate the accuracy of the current method for determining the
structural parameters of these RP phases, we also examined the structural
properties of the parent materials, specifically bulk AHfS$_{3}$
($n$ = $\infty$), where A = Ca, Sr, and Ba. It is found that the
relaxed lattice parameters of these parent materials are in excellent
agreement with the experimental values \citep{chapter3-37}. For example,
the experimental lattice parameters of SrHfS$_{3}$ are a = 6.72 $\textrm{\AA}$,
b = 7.05 $\textrm{\AA}$, and c = 9.72 $\textrm{\AA}$, which closely
align with our theoretical values. This agreement validates the reliability
of the PBE method in accurately describing the structural properties
of RP perovskites.

\begin{table}[H]
\caption{\label{Table:1}Calculated lattice parameters a, b, and c (in unit
$\textrm{\AA}$) of A$_{n+1}$Hf$_{n}$S$_{3n+1}$ (A = Ca, Sr, Ba;
$n$ = 1$-$3) RP phases and AHfS$_{3}$ (A = Ca, Sr, Ba; $n$ = $\infty$)
chalcogenide perovskites.}

\centering{}%
\begin{tabular}{ccccccccc}
\hline 
Configurations &  & $n$ = 1 &  & $n$ = 2 &  & $n$ = 3 &  & $n$ = $\infty$\tabularnewline
\hline 
\multirow{2}{*}{Ca$_{n+1}$Hf$_{n}$S$_{3n+1}$} &  & a = b = 4.90, &  & a = b = 4.93, &  & a = b = 4.94, &  & a = 6.56, b = 7.01,\tabularnewline
 &  & c = 14.36 &  & c = 23.85 &  & c = 33.78 &  & c = 9.58\tabularnewline
\multirow{2}{*}{Sr$_{n+1}$Hf$_{n}$S$_{3n+1}$} &  & a = b = 4.91, &  & a = b = 4.95, &  & a = b = 4.97, &  & a = 6.76, b = 7.10,\tabularnewline
 &  & c = 15.13 &  & c = 24.81 &  & c = 34.67 &  & c = 9.77\tabularnewline
\multirow{2}{*}{Ba$_{n+1}$Hf$_{n}$S$_{3n+1}$} &  & a = b = 4.96, &  & a = b = 4.99, &  & a = b = 5.03, &  & a = 7.03, b = 7.09,\tabularnewline
 &  & c = 15.86 &  & c = 25.75 &  & c = 35.63 &  & c = 9.99\tabularnewline
\hline 
\end{tabular}
\end{table}

To assess the stability of these RP phases, their mechanical stability
and associated elastic properties have also been analyzed. The elastic
properties are computed by determining the second-order elastic constants
($C_{ij}$) using the energy-strain approach \citep{chapter1-47}.
Since all the RP phases considered in this study exhibit tetragonal
(I) symmetry, six independent elastic coefficients\textemdash $C_{11}$,
$C_{33}$, $C_{44}$, $C_{66}$, $C_{12}$, and $C_{13}$\textemdash are
required to explain the mechanical stability and corresponding elastic
properties of these systems. Using these elastic constants, we can
calculate the bulk modulus ($B$), shear modulus ($G$), Young\textquoteright s
modulus ($Y$), and Poisson\textquoteright s ratio ($\nu$) of the
perovskites (for details, see the Supplemental Material). The corresponding
mechanical stability criterion is expressed as follows \citep{chapter1-47}:

\begin{equation}
C_{11}>|C_{12}|,\;2C_{13}^{2}<C_{33}(C_{11}+C_{12}),\;C_{44}>0,\;C_{66}>0
\end{equation}

The calculated elastic constants and moduli are provided in Table
\ref{Table:2}. As shown, the elastic constants meet the Born stability
criteria, confirming the mechanical stability of these RP phases.
Pugh\textquoteright s suggested $B/G$ ratio and Poisson\textquoteright s
ratio ($\nu$) are used to study the fragility of the materials \citep{chapter1-51}.
A material is considered ductile if $B/G$ is greater than 1.75 and
($\nu$) exceeds 0.26; otherwise, it is brittle. The calculated values
of $B/G$ and ($\nu$) indicate that the studied compounds exhibit
ductile behavior. These specific elastic properties make the selected
RP phases ideal for applications in flexible and durable devices.

\begin{table}[H]
\caption{\label{Table:2}Mechanical parameters for RP phases, including elastic
constants $C_{ij}$ (in GPa), bulk modulus $B$ (in GPa), shear modulus
$G$ (in GPa), Young\textquoteright s modulus $Y$ (in GPa), and Poisson\textquoteright s
ratio $\nu$ (dimensionless).}

\centering{}%
\begin{tabular}{ccccccccccccc}
\hline 
\multicolumn{2}{c}{Configurations} & $C_{11}$ & $C_{33}$ & $C_{44}$ & $C_{66}$ & $C_{12}$ & $C_{13}$ & $B$ & $G$ & $Y$ & $B/G$ & $\nu$\tabularnewline
\hline 
\multirow{3}{*}{Ca$_{n+1}$Hf$_{n}$S$_{3n+1}$} & $n$ = 1 & 130.20 & 121.70 & 24.00 & 15.41 & 26.63 & 41.36 & 66.74 & 28.38 & 74.56 & 2.35 & 0.31\tabularnewline
 & $n$ = 2 & 133.24 & 112.47 & 20.02 & 21.92 & 30.00 & 37.76 & 65.47 & 28.45 & 74.55 & 2.30 & 0.31\tabularnewline
 & $n$ = 3 & 125.91 & 93.18 & 17.46 & 29.55 & 29.26 & 39.02 & 61.86 & 26.76 & 70.16 & 2.31 & 0.31\tabularnewline
\multirow{3}{*}{Sr$_{n+1}$Hf$_{n}$S$_{3n+1}$} & $n$ = 1 & 130.03 & 116.23 & 23.98 & 19.61 & 25.25 & 37.45 & 64.06 & 29.95 & 77.74 & 2.14 & 0.30\tabularnewline
 & $n$ = 2 & 133.92 & 112.19 & 24.04 & 24.30 & 28.09 & 37.86 & 65.23 & 31.25 & 80.85 & 2.09 & 0.29\tabularnewline
 & $n$ = 3 & 125.15 & 110.56 & 21.39 & 30.35 & 31.33 & 39.43 & 64.56 & 29.95 & 77.82 & 2.16 & 0.30\tabularnewline
\multirow{3}{*}{Ba$_{n+1}$Hf$_{n}$S$_{3n+1}$} & $n$ = 1 & 121.09 & 107.77 & 25.46 & 24.72 & 24.94 & 31.09 & 58.22 & 31.46 & 79.98 & 1.85 & 0.27\tabularnewline
 & $n$ = 2 & 132.43 & 116.91 & 29.21 & 28.53 & 29.64 & 35.38 & 64.69 & 35.13 & 89.23 & 1.84 & 0.27\tabularnewline
 & $n$ = 3 & 123.13 & 114.50 & 23.51 & 34.74 & 32.44 & 34.67 & 62.68 & 32.40 & 82.92 & 1.93 & 0.28\tabularnewline
\hline 
\end{tabular}
\end{table}

\subsection{Electronic properties:}

\begin{figure}[H]
\begin{centering}
\includegraphics[width=1\textwidth,height=1\paperheight,keepaspectratio]{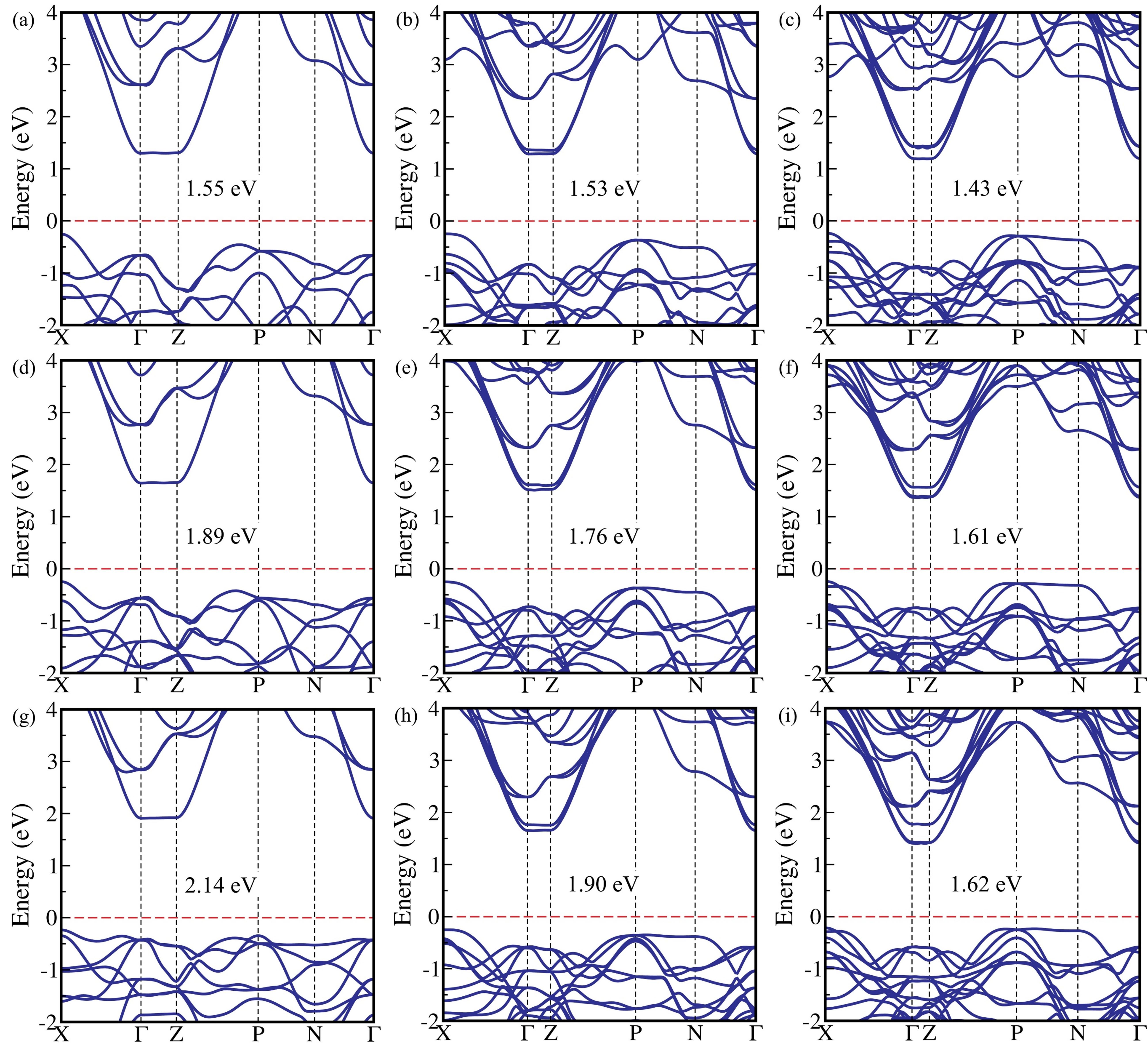}
\par\end{centering}
\caption{\label{Figure:2}Band structures of (a$-$c) Ca$_{n+1}$Hf$_{n}$S$_{3n+1}$,
(d\textminus f) Sr$_{n+1}$Hf$_{n}$S$_{3n+1}$, and (g\textminus i)
Ba$_{n+1}$Hf$_{n}$S$_{3n+1}$, respectively, where $n$ = 1$-$3,
calculated using the G$_{0}$W$_{0}$@PBE method.}
\end{figure}

After evaluating the stability of the A$_{n+1}$Hf$_{n}$S$_{3n+1}$
(A = Ca, Sr, Ba; $n$ = 1$-$3) RP phases, their electronic band structures
are calculated, which play a crucial role in the design of photoelectric
devices. Initially, the band structures are computed using the semi-local
PBE xc functional, both with and without considering the spin-orbit
coupling (SOC) effect. However, it is well-known that the PBE functional
struggles to accurately predict the bandgaps of chalcogenide perovskites
due to its inability to account for the self-interaction error of
electrons \citep{chapter3-19,chapter5-16}. Additionally, the inclusion
of SOC has no significant impact on the bandgaps (For details, see
the Table S1). To achieve more accurate results, hybrid HSE06 functional
(with 25\% exact exchange and a screening parameter of 0.2 $\textrm{\AA}^{-1}$)
and the MBPT-based G$_{0}$W$_{0}$@PBE method are employed for bandgap
calculations. The band structures calculated using G$_{0}$W$_{0}$@PBE
method are presented in Figure \ref{Figure:2}, while those obtained
with the HSE06 functional are shown in Figure S1. Our results show
that all the RP phases display indirect bandgaps, with the conduction
band minimum (CBM) located at the $\Gamma$ (0, 0, 0) point and the
valence band maximum (VBM) situated at the X (0, 0, 0.5) point of
the Brillouin zone. The HSE06 and G$_{0}$W$_{0}$@PBE estimated bandgaps
of these RP phases are listed in Table \ref{Table:3}, showing that
the bandgaps of these compounds decrease as the thickness of the perovskite
layer ($n$) increases. The bandgap of these RP phases, computed using
HSE06 and G$_{0}$W$_{0}$@PBE, lies within the ranges of 1.02$-$1.66
eV and 1.43$-$2.14 eV, respectively. This implies that all of these
RP phases have smaller bandgaps compared to their bulk counterparts,
such as AHfS$_{3}$ ($n=\infty$, where A = Ca, Sr, Ba) \citep{chapter3-19,chapter7-2}.
For example, CaHfS$_{3}$ perovskite has a G$_{0}$W$_{0}$@PBE bandgap
of 2.46 eV, which is significantly higher than the bandgaps of the
Ca$_{n+1}$Hf$_{n}$S$_{3n+1}$ ($n$ = 1$-$3) RP phases. This property
suggests that these RP phases could be promising candidates for optoelectronic
technologies.

\begin{table}[H]
\caption{\label{Table:3}Bandgap (in eV) of RP phases calculated using PBE,
HSE06 and G$_{0}$W$_{0}$@PBE method. Also, effective masses of electron
($m_{e}^{*}$), hole ($m_{h}^{*}$), and reduced mass ($\mu^{*}$)
of RP phases calculated using G$_{0}$W$_{0}$@PBE method.}

\centering{}%
\begin{tabular}{ccccccccc}
\hline 
\multicolumn{2}{c}{Configurations} &  & PBE & HSE06 & G$_{0}$W$_{0}$@PBE & $m_{e}^{*}$ ($m_{0}$) & $m_{h}^{*}$ ($m_{0}$) & $\mu^{*}$ ($m_{0}$)\tabularnewline
\hline 
\multirow{3}{*}{Ca$_{n+1}$Hf$_{n}$S$_{3n+1}$} & $n$ = 1 &  & 0.32 & 1.08 & 1.55 & 0.179 & 0.573 & 0.136\tabularnewline
 & $n$ = 2 &  & 0.30 & 1.05 & 1.53 & 0.182 & 1.683 & 0.164\tabularnewline
 & $n$ = 3 &  & 0.24 & 1.02 & 1.43 & 0.192 & 0.537 & 0.141\tabularnewline
\multirow{3}{*}{Sr$_{n+1}$Hf$_{n}$S$_{3n+1}$} & $n$ = 1 &  & 0.61 & 1.38 & 1.89 & 0.177 & 0.726 & 0.142\tabularnewline
 & $n$ = 2 &  & 0.55 & 1.31 & 1.76 & 0.179 & 2.365 & 0.166\tabularnewline
 & $n$ = 3 &  & 0.46 & 1.28 & 1.61 & 0.187 & 0.597 & 0.142\tabularnewline
\multirow{3}{*}{Ba$_{n+1}$Hf$_{n}$S$_{3n+1}$} & $n$ = 1 &  & 0.87 & 1.66 & 2.14 & 0.185 & 0.868 & 0.152\tabularnewline
 & $n$ = 2 &  & 0.75 & 1.52 & 1.90 & 0.183 & 3.880 & 0.175\tabularnewline
 & $n$ = 3 &  & 0.55 & 1.37 & 1.62 & 0.192 & 0.600 & 0.145\tabularnewline
\hline 
\end{tabular}
\end{table}

To gain deeper insight into charge carrier transport, we further calculated
the effective masses of electrons ($m_{e}^{*}$) and holes ($m_{h}^{*}$)
for all the RP phases. This is achieved by fitting the $E-k$ dispersion
from the G$_{0}$W$_{0}$@PBE band structures and applying the formula:
$\text{\ensuremath{m^{*}=\hbar^{2}\left[\partial^{2}E(k)/\partial k^{2}\right]^{-1}}}$,
where $\hbar$ is the reduced Planck's constant. In this study, we
computed the effective masses of electrons and holes along the symmetric
paths $\Gamma\to\mathrm{X}$ and $\mathrm{X}\to\Gamma$, respectively.
The obtained values of the effective mass and reduced mass, expressed
in terms of the rest mass of an electron ($m_{0}$), for the RP phases
are listed in Table \ref{Table:3}. Also, it is found that the electron
effective masses of these RP phases increase with the value of $n$,
whereas the hole effective masses exhibit an oscillating behavior.
Overall, these results suggest that the RP phases are likely to exhibit
higher charge carrier mobility for electrons compared to holes.

\begin{figure}[H]
\begin{centering}
\includegraphics[width=1\textwidth,height=1\textheight,keepaspectratio]{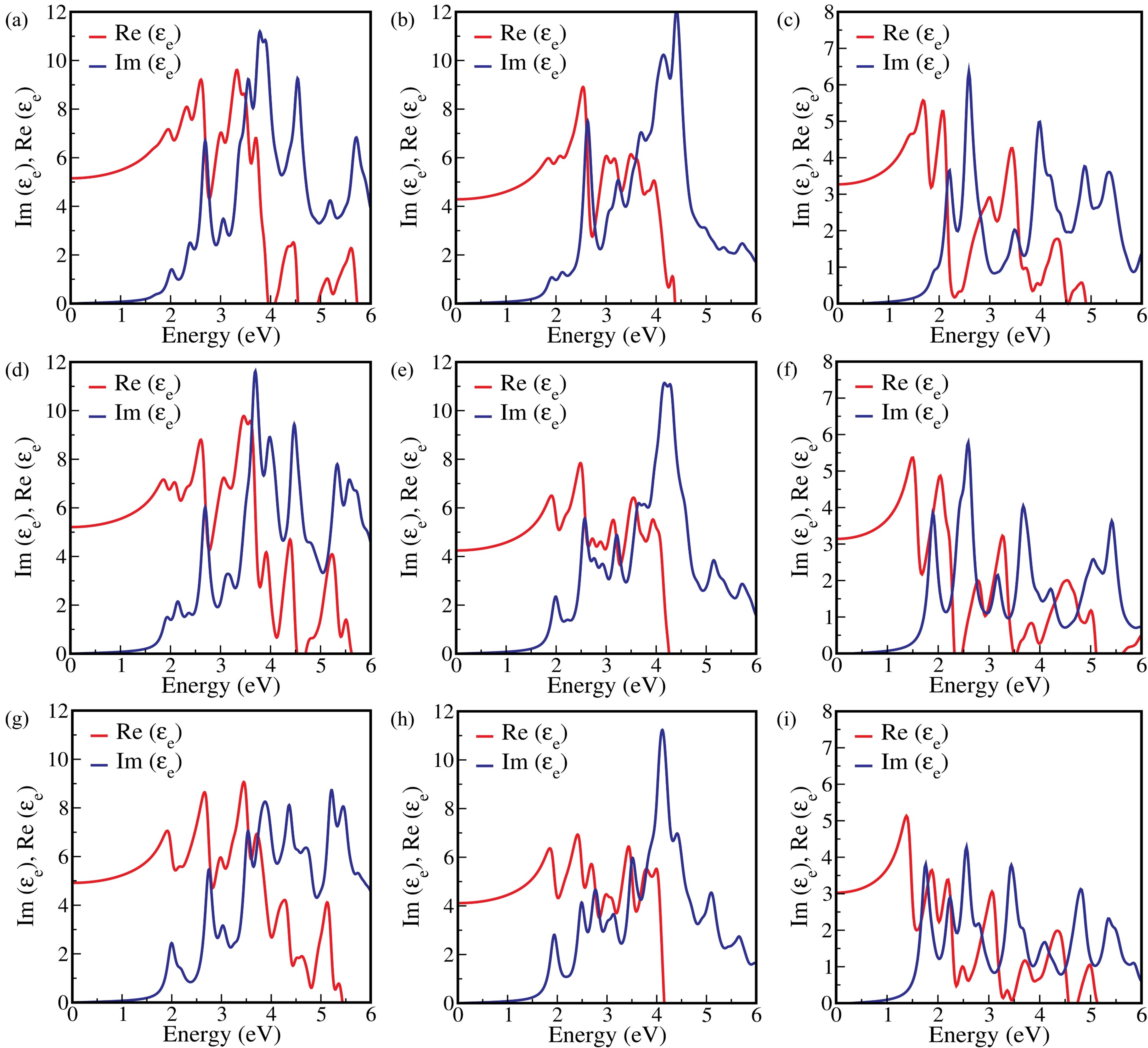}
\par\end{centering}
\caption{\label{Figure:3}Real {[}Re($\varepsilon$){]} and imaginary {[}Im($\varepsilon$){]}
parts of the dielectric (electronic) function for (a$-$c) Ca$_{n+1}$Hf$_{n}$S$_{3n+1}$,
(d$-$f) Sr$_{n+1}$Hf$_{n}$S$_{3n+1}$, and (g$-$i) Ba$_{n+1}$Hf$_{n}$S$_{3n+1}$,where
$n$ = 1\textminus 3, computed using the BSE@G$_{0}$W$_{0}$@PBE
($n$ = 1, 2) and mBSE@G$_{0}$W$_{0}$@PBE ($n$ = 3) methods, respectively.}
\end{figure}

\subsection{Optical Properties:}

Although most of these compounds exhibit lower bandgaps and low effective
carrier masses, these properties alone are insufficient to guarantee
their suitability for optoelectronic device applications. Therefore,
we investigated the optical response of the A$_{n+1}$Hf$_{n}$S$_{3n+1}$
(A = Ca, Sr, Ba; $n$ = 1$-$3) RP phases by calculating the real
{[}Re($\varepsilon$){]} and imaginary {[}Im($\varepsilon$){]} parts
of the dielectric function. To achieve higher reliability in optical
response estimation, the Bethe-Salpeter Equation (BSE) was solved
on top of G$_{0}$W$_{0}$@PBE, explicitly accounting for excitonic
effects (electron-hole interactions) \citep{chapter1-67,chapter1-68}.
However, for $n$ = 3 in the RP phases, performing BSE calculations
has been computationally infeasible, even with the most advanced supercomputers.
To address this limitation, a parametrized model for dielectric screening,
known as the model-BSE (mBSE) approach \citep{chapter1-65}, has been
employed for these compounds. This mBSE approach is computationally
less expensive compared to the full BSE, while maintaining similar
accuracy in estimating the first peak position (for details, see the
Supplemental Material).

Figure \ref{Figure:3} represents the average optical responses, including
the real and imaginary components of the electronic dielectric function,
calculated along the $x$, $y$, and $z$ directions for these RP
phases. However, Figure S3{[}(a)-(c){]} and Figure S3 {[}(d)-(f){]}
illustrate the directional optical responses of the RP phases along
the E$\bigparallel$$xy$ (i.e., in-plane along the $x$ or $y$ directions)
and E$\bigparallel$$z$ (i.e.,out-of-plane along the $z$ direction)
directions, respectively. The directional optical responses exhibit
significant optical anisotropy in these compounds, which could have
a substantial impact on their performance in practical applications.
Figure \ref{Figure:3} reveals that the absorption onset of these
compounds shows a gradual redshift with an increase in the thickness
of the perovskite layers. Additionally, the optical peak position
($E_{o}$) of these compounds varies with the $n$ value, which is
crucial for determining the exciton binding energy. For example, the
optical peak position of Ba$_{2}$HfS$_{4}$ is 2 eV, while those
of Ba$_{3}$Hf$_{2}$S$_{7}$ and Ba$_{4}$Hf$_{3}$S$_{10}$ are
1.94 eV and 1.76 eV, respectively. Similarly, the peak positions of
Ca$_{n+1}$Hf$_{n}$S$_{3n+1}$ and Sr$_{n+1}$Hf$_{n}$S$_{3n+1}$
varies in between 1.72$-$1.91 eV and 1.89$-$1.99 eV, respectively.
It should be emphasized that all the compounds exhibit absorption
onsets within the visible region, making them well-suited for optoelectronic
applications.

\subsection{Excitonic Properties:}

In optoelectronic materials, exciton generation significantly influences
charge separation properties. Consequently, excitonic parameters,
such as exciton binding energy ($E_{B}$) and exciton lifetime ($\tau_{exc}$),
serve as critical descriptors for evaluating and optimizing their
performance in various applications. The exciton binding energy ($E_{B}$)
is defined as the energy required to dissociate an exciton into its
constituent electron ($e$) and hole ($h$) pair. The value of $E_{B}$
is calculated using the Wannier$\lyxmathsym{\textendash}$Mott model,
which assumes a simple screened Coulomb potential to describe the
interaction between the electron and hole in a material \citep{Intro25}.
Recent studies have reported that excitons in the RP phases of MAPbX$_{3}$
are significantly influenced by quantum confinement effects along
the $z$-direction, leading to a quasi-2D character \citep{chapter7-3}.
In our case, we observe differences in the spectra along the $xy$
and $z$ directions, indicating weak hybridization along the $z$-direction.
Also, increasing the number of perovskite layers leads to deviations
from ideal 2D behavior, attributed to the screened Coulomb potential
\citep{chapter7-3,Intro25}. Therefore, For $n$ = 1 and 2, the 2D
Wannier\textendash Mott model is more applicable, while for $n$ =
3 and above, the 3D model may also be relevant. Thus, according to
the 2D Wannier\textminus Mott model, the $E_{B}$ of the RP phases
is calculated as \citep{chapter7-3,Intro25},

\begin{equation}
E_{B}=\frac{1}{\left(n-\frac{1}{2}\right)^{2}}\left(\frac{\mu^{*}}{\mathrm{\varepsilon_{eff}^{2}}}\right)R_{\infty}\label{eq:2}
\end{equation}

where $\mu^{\ast}$ is the reduced mass of the charge carriers at
the direct band edges, $m_{0}$ is the rest mass of the electron,
$\varepsilon_{\mathrm{eff}}$ denotes the effective dielectric constant,
and $R_{\infty}$ is the Rydberg constant. The reduced masses for
the RP phases at direct band edges are provided in Table S2 of the
Supplemental Material. However, $\varepsilon_{\mathrm{eff}}$ is unknown
for these RP phases. According to previous reports, $E_{B}$ is influenced
by lattice relaxation \citep{chapter1-64}, and if $E_{B}\ll\hbar\omega_{LO}$,
lattice relaxation must be considered, where $\omega_{LO}$ denotes
the longitudinal optical phonon frequency. Therefore, in determining
$\varepsilon_{\mathrm{eff}}$, a value intermediate between the electronic
$(\ensuremath{\varepsilon_{\infty}})$ and static ($\varepsilon_{s}=\varepsilon_{\infty}+\varepsilon_{i}$)
dielectric constants should be considered, where $\varepsilon_{i}$
representing the ionic contribution to the dielectric function. The
values of $\varepsilon_{\infty}$ and $\varepsilon_{i}$ are obtained
using the BSE/mBSE and DFPT methods, respectively. On the other hand,
if $E_{B}\gg\hbar\omega_{LO}$, the ionic dielectric constant becomes
negligible, and $\varepsilon_{\mathrm{eff}}\rightarrow\varepsilon_{\infty}$,
indicating that the static dielectric constant is primarily determined
by the electronic contribution at high frequency \citep{chapter1-65,chapter1-66}.
Using the electronic and static dielectric constants, along with Equation
\ref{eq:2}, we have calculated the upper ($E_{Bu}$) and lower ($E_{Bl}$)
bounds of the exciton binding energy (see Table \ref{Table:4}). The
results indicate that the upper bounds of $E_{B}$ show a significant
reduction in exciton binding energies for the examined RP phases as
the thickness of the perovskite layer ($n$) increases. In the absence
of experimental data, we have also conducted a theoretical analysis
based on the 3D Wannier\textendash Mott excitons (for details, see
the Supplemental Material).

\begin{table}[H]
\caption{\label{Table:4}Exciton binding energy of A$_{n+1}$Hf$_{n}$S$_{3n+1}$
(A = Ca, Sr, Ba and $n$ = 1$-$3) RP phases using Wannier-Mott Model.}

\centering{}%
\begin{tabular}{cccccc}
\hline 
\multicolumn{2}{c}{Configurations} & $\varepsilon_{\infty}$ & $E_{Bu}$ (meV) & $\varepsilon_{s}$ & $E_{Bl}$ (meV)\tabularnewline
\hline 
\multirow{3}{*}{Ca$_{n+1}$Hf$_{n}$S$_{3n+1}$} & $n$ = 1 & 5.46 & 218.98 & 17.51 & 21.29\tabularnewline
 & $n$ = 2 & 4.35 & 38.97 & 22.71 & 1.43\tabularnewline
 & $n$ = 3 & 2.03 & 89.23 & 22.59 & 0.72\tabularnewline
\multirow{3}{*}{Sr$_{n+1}$Hf$_{n}$S$_{3n+1}$} & $n$ = 1 & 5.62 & 218.74 & 42.22 & 3.88\tabularnewline
 & $n$ = 2 & 4.42 & 40.22 & 136.58 & 0.04\tabularnewline
 & $n$ = 3 & 2.06 & 74.35 & 17.78 & 1.00\tabularnewline
\multirow{3}{*}{Ba$_{n+1}$Hf$_{n}$S$_{3n+1}$} & $n$ = 1 & 5.22 & 309.45 & 47.68 & 3.71\tabularnewline
 & $n$ = 2 & 4.34 & 51.99 & 92.70 & 0.11\tabularnewline
 & $n$ = 3 & 2.15 & 77.20 & 274.79 & 0.005\tabularnewline
\hline 
\end{tabular}
\end{table}

In addition to the Wannier-Mott model, we computed the first-principles
exciton binding energies ($E_{B}$) for these RP phases using the
BSE (for $n$ = 1, 2) and mBSE (for $n$ = 3) methods. The binding
energy is defined as, $E_{B}$ = $E_{g}^{dir}$ $-$ $E_{o}$, where
$E_{g}^{dir}$ is the direct G$_{0}$W$_{0}$@PBE bandgap, and $E_{o}$
represents the position of the first prominent absorption peak obtained
from the BSE/mBSE calculations \citep{chapter2-16,chapter5-18}. Table
\ref{Table:5} indicates that the $E_{B}$ values of these RP phases
range from 0.17 to 0.33 eV, decreasing with increasing $n$. This
trend aligns with the predictions of the Wannier-Mott model. Tables
\ref{Table:5} and \ref{Table:6} show that $E_{B}\gg\hbar\omega_{LO}$,
indicating that the electronic contribution dominates over the ionic
contribution in dielectric screening ($\varepsilon_{\mathrm{eff}}\rightarrow\varepsilon_{\infty}$)
for RP phases. The studies also reveal that the $E_{B}$ values of
these RP phases are generally smaller compared to their bulk counterparts
($n$ = $\infty$), with the exception of Ba-based compounds \citep{chapter3-19}.
In view of this, the considered RP phases are anticipated to be superior
optoelectronic materials compared to their bulk counterparts.

\begin{table}[H]
\caption{\label{Table:5}Excitonic parameters for A$_{n+1}$Hf$_{n}$S$_{3n+1}$
(A = Ca, Sr, Ba and $n$ = 1$-$3) RP phases using BSE/mBSE approach.}

\centering{}%
\begin{tabular}{ccccccc}
\hline 
\multicolumn{2}{c}{Configurations} & $E_{B}$ (eV) & $T_{exc}$ (K) & $r_{exc}$ (nm) & $|\phi_{l}(0)|^{2}$ ($10^{27}$$m^{-3}$) & $\frac{1}{|\phi_{l}(0)|^{2}}$ ($10^{-27}$$m^{3}$)\tabularnewline
\hline 
\multirow{3}{*}{Ca$_{n+1}$Hf$_{n}$S$_{3n+1}$} & $n$ = 1 & 0.24 & 2783 & 1.20 & 0.18 & 5.48\tabularnewline
 & $n$ = 2 & 0.21 & 2435 & 0.94 & 0.38 & 2.64\tabularnewline
 & $n$ = 3 & 0.17 & 1971 & 0.32 & 9.93 & 0.10\tabularnewline
\multirow{3}{*}{Sr$_{n+1}$Hf$_{n}$S$_{3n+1}$} & $n$ = 1 & 0.27 & 3130 & 1.17 & 0.20 & 5.04\tabularnewline
 & $n$ = 2 & 0.26 & 3014 & 0.90 & 0.44 & 2.28\tabularnewline
 & $n$ = 3 & 0.23 & 2667 & 0.38 & 6.00 & 0.17\tabularnewline
\multirow{3}{*}{Ba$_{n+1}$Hf$_{n}$S$_{3n+1}$} & $n$ = 1 & 0.33 & 3826 & 0.89 & 0.45 & 2.22\tabularnewline
 & $n$ = 2 & 0.29 & 3362 & 0.71 & 0.89 & 1.12\tabularnewline
 & $n$ = 3 & 0.23 & 2667 & 0.35 & 7.63 & 0.13\tabularnewline
\hline 
\end{tabular}
\end{table}

Next, additional excitonic parameters, including the excitonic temperature
($T_{exc}$), radius ($r_{exc}$), and probability of wave function
($|\phi_{l}(0)|^{2}$) for $e-h$ pair at zero separation, have been
calculated (for details, see the Supplemental Material). The computed
values of these parameters are presented in Table \ref{Table:5}.
It is well established that the inverse of $|\phi_{l}(0)|^{2}$ serves
as a qualitative indicator of exciton lifetime ($\tau_{exc}$; for
details, see the Supplemental Material). Table \ref{Table:5} suggests
that the value of $|\phi_{l}(0)|^{2}$ increases with increasing $n$
in these RP phases. Consequently, the exciton lifetime ($\tau_{exc}$)
is expected to decrease with higher values of $n$ in these phases.
A longer exciton lifetime corresponds to a reduced carrier recombination
rate, thereby enhancing the quantum yield and improving the conversion
efficiency.

\subsection{Polaronic Properties:}

It is well known that polar materials frequently undergo polaron formation,
which significantly affects the mobility of charge carriers. The Fr\"ohlich
model provides a comprehensive description of the interaction between
charge carriers and polar lattice vibrations \citep{chapter2-51}.
Fr\"ohlich proposed a dimensionless descriptor, $\alpha$, to quantify
the strength of this interaction, as follows \citep{chapter5-16}:

\begin{equation}
\alpha=\frac{1}{4\pi\varepsilon_{0}}\frac{1}{2}\Big(\frac{1}{\varepsilon_{\infty}}-\frac{1}{\varepsilon_{s}}\Big)\frac{e^{2}}{\hbar\omega_{LO}}\Big(\frac{2m^{\ast}\omega_{LO}}{\hbar}\Big)^{1/2}
\end{equation}

where $\varepsilon_{0}$ is the permittivity of free space, $m^{\ast}$
represents the effective mass of the carrier, and $\omega_{LO}$ is
the characteristic phonon angular frequency. The thermal \textquotedbl B\textquotedbl{}
approach, developed by Hellwarth et al. \citep{chapter2-22}, is used
to determine $\omega_{LO}$ by averaging the spectral contributions
of multiple phonon branches (for details, see the Supplemental Material).
The computed carrier-phonon coupling constants ($\alpha$) for both
electrons and holes in these RP phases are listed in Table \ref{Table:6}.
It is observed that the considered RP phases exhibit intermediate
to strong carrier-phonon coupling, which is significant and cannot
be neglected for these materials \citep{chapter2-20}. The Fr\"ohlich
model also reveals that, for these compounds, the hole-phonon coupling
is significantly stronger than the electron-phonon coupling. Additionally,
the Debye temperature ($\theta_{D}$) for these RP phases is calculated
and found to be lower than room temperature (except Ca$_{2}$HfS$_{4}$),
suggesting a strong interaction between carriers and phonons.

Polaron formation notably causes a reduction in the quasiparticle
(QP) energies of both electrons and holes. This polaron energy ($E_{p}$)
can also be determined using the parameter $\alpha$, as follows \citep{chapter5-16}:

\begin{equation}
E_{p}=(-\alpha-0.0123\alpha^{2})\hbar\omega_{LO}
\end{equation}

Table \ref{Table:6} indicates that the energy of the charge-separated
polaronic states in these RP phases is lower than that of their bound
exciton states. This implies that the charge-separated polaronic states
are less stable compared to the bound excitons.

Further, based on the extended form of Fr\"ohlich\textquoteright s polaron
theory, as developed by Feynman, the effective mass of the polaron
($\ensuremath{m_{p}}$) is expressed as (for a small $\alpha$) \citep{chapter2-23}:

\begin{equation}
m_{p}=m^{\ast}\Big(1+\frac{\alpha}{6}+\frac{\alpha^{2}}{40}+...\Big)
\end{equation}

The effective masses of electrons and holes increase due to polaron
formation, with the extent of this increase quantified to range between
27\% and 644\% in these materials (see Table \ref{Table:6}). These
results demonstrate that electron-phonon coupling leads to a significant
increase in the polaron effective mass, suggesting intermediate to
strong carrier-lattice interactions. In addition, the upper bound
of the charge-carrier mobilities ($\mu_{p}$) for these RP phases
is estimated using Feynman\textquoteright s variational solution to
Fr\"ohlich\textquoteright s polaron Hamiltonian and by integrating the
polaron response function \citep{chapter2-22,chapter2-20,chapter2-23}
(for details, see the Supplemental Material). The polaron mobility,
as described by the Hellwarth polaron model, is defined as follows
\citep{chapter2-22}:

\begin{equation}
\mu_{p}=\frac{\left(3\sqrt{\pi}e\right)}{2\pi c\omega_{LO}m^{*}\alpha}\frac{\sinh(\beta/2)}{\beta^{5/2}}\frac{w^{3}}{v^{3}}\frac{1}{K(a,b)}
\end{equation}
where $e$ denontes the electronic charge, $\beta=hc\omega_{LO}/k_{B}T$,
$w$ and $v$ both are the temperature-dependent variational parameters,
and $K(a,b)$ is a function of $\beta$, $w$, and $v$ (for details,
see the Supplemental Material). Our findings (see Table \ref{Table:6})
reveal that as $n$ increases in the considered RP phases, the polaron
mobility of electrons decreases, becoming very small in their bulk
counterparts ($n$ = $\infty$) \citep{chapter3-19}. In contrast,
the polaron mobility of holes displays an oscillating behavior. The
polaron mobility values span from 26.32 to 121.32 cm$^{2}$V$^{-1}$s$^{-1}$
for electrons and from 0.04 to 17.14 cm$^{2}$V$^{-1}$s$^{-1}$ for
holes. In view of this, the considered RP phases are expected to be
superior optoelectronic materials compared to their bulk counterparts.

\begin{table}[H]
\caption{\label{Table:6}Polaron parameters for electrons ($e$) and holes
($h$) of A$_{n+1}$Hf$_{n}$S$_{3n+1}$ (A = Ca, Sr, Ba and $n$
= 1$-$3) RP phases.}

\centering{}{\scriptsize{}}%
\begin{tabular}{ccccccccccccccc}
\hline 
\multicolumn{2}{c}{} & \multirow{2}{*}{{\scriptsize{}$\omega_{LO}$ (THz)}} & \multirow{2}{*}{{\scriptsize{}$\theta_{D}$ (K)}} & \multicolumn{2}{c}{{\scriptsize{}$\alpha$}} &  & \multicolumn{2}{c}{{\scriptsize{}$E_{p}$ (meV)}} &  & \multicolumn{2}{c}{{\scriptsize{}$m_{p}/m^{*}$}} &  & \multicolumn{2}{c}{{\scriptsize{}$\mu_{p}$ (cm$^{2}$V$^{-1}$s$^{-1}$)}}\tabularnewline
\cline{5-6} \cline{6-6} \cline{8-9} \cline{9-9} \cline{11-12} \cline{12-12} \cline{14-15} \cline{15-15} 
\multicolumn{2}{c}{{\scriptsize{}Configurations}} &  &  & {\scriptsize{}$e$} & {\scriptsize{}$h$} &  & {\scriptsize{}$e$} & {\scriptsize{}$h$} &  & {\scriptsize{}$e$} & {\scriptsize{}$h$} &  & {\scriptsize{}$e$} & {\scriptsize{}$h$}\tabularnewline
\hline 
\multirow{3}{*}{{\scriptsize{}Ca$_{n+1}$Hf$_{n}$S$_{3n+1}$}} & {\scriptsize{}$n$ = 1} & {\scriptsize{}6.51} & {\scriptsize{}312.57} & {\scriptsize{}1.35} & {\scriptsize{}2.41} &  & {\scriptsize{}37.00} & {\scriptsize{}66.90} &  & {\scriptsize{}1.27} & {\scriptsize{}1.55} &  & {\scriptsize{}121.32} & {\scriptsize{}17.14}\tabularnewline
 & {\scriptsize{}$n$ = 2} & {\scriptsize{}3.63} & {\scriptsize{}174.29} & {\scriptsize{}2.69} & {\scriptsize{}8.17} &  & {\scriptsize{}41.78} & {\scriptsize{}135.16} &  & {\scriptsize{}1.63} & {\scriptsize{}4.03} &  & {\scriptsize{}58.81} & {\scriptsize{}0.65}\tabularnewline
 & {\scriptsize{}$n$ = 3} & {\scriptsize{}3.58} & {\scriptsize{}171.89} & {\scriptsize{}3.52} & {\scriptsize{}5.89} &  & {\scriptsize{}54.45} & {\scriptsize{}93.65} &  & {\scriptsize{}1.90} & {\scriptsize{}2.85} &  & {\scriptsize{}37.21} & {\scriptsize{}4.97}\tabularnewline
\multirow{3}{*}{{\scriptsize{}Sr$_{n+1}$Hf$_{n}$S$_{3n+1}$}} & {\scriptsize{}$n$ = 1} & {\scriptsize{}4.72} & {\scriptsize{}226.63} & {\scriptsize{}1.80} & {\scriptsize{}3.65} &  & {\scriptsize{}35.96} & {\scriptsize{}74.55} &  & {\scriptsize{}1.38} & {\scriptsize{}1.94} &  & {\scriptsize{}93.25} & {\scriptsize{}7.86}\tabularnewline
 & {\scriptsize{}$n$ = 2} & {\scriptsize{}4.14} & {\scriptsize{}198.78} & {\scriptsize{}2.68} & {\scriptsize{}9.76} &  & {\scriptsize{}47.46} & {\scriptsize{}187.42} &  & {\scriptsize{}1.63} & {\scriptsize{}5.01} &  & {\scriptsize{}56.18} & {\scriptsize{}0.21}\tabularnewline
 & {\scriptsize{}$n$ = 3} & {\scriptsize{}1.53} & {\scriptsize{}73.46} & {\scriptsize{}5.50} & {\scriptsize{}9.83} &  & {\scriptsize{}37.21} & {\scriptsize{}69.81} &  & {\scriptsize{}2.67} & {\scriptsize{}5.05} &  & {\scriptsize{}33.52} & {\scriptsize{}3.19}\tabularnewline
\multirow{3}{*}{{\scriptsize{}Ba$_{n+1}$Hf$_{n}$S$_{3n+1}$}} & {\scriptsize{}$n$ = 1} & {\scriptsize{}4.47} & {\scriptsize{}214.62} & {\scriptsize{}2.06} & {\scriptsize{}4.45} &  & {\scriptsize{}39.10} & {\scriptsize{}86.88} &  & {\scriptsize{}1.45} & {\scriptsize{}2.24} &  & {\scriptsize{}76.15} & {\scriptsize{}4.69}\tabularnewline
 & {\scriptsize{}$n$ = 2} & {\scriptsize{}3.91} & {\scriptsize{}187.74} & {\scriptsize{}2.84} & {\scriptsize{}13.06} &  & {\scriptsize{}47.59} & {\scriptsize{}245.44} &  & {\scriptsize{}1.67} & {\scriptsize{}7.44} &  & {\scriptsize{}51.91} & {\scriptsize{}0.04}\tabularnewline
 & {\scriptsize{}$n$ = 3} & {\scriptsize{}3.59} & {\scriptsize{}172.37} & {\scriptsize{}4.30} & {\scriptsize{}7.61} &  & {\scriptsize{}67.31} & {\scriptsize{}123.73} &  & {\scriptsize{}2.18} & {\scriptsize{}3.72} &  & {\scriptsize{}26.32} & {\scriptsize{}2.29}\tabularnewline
\hline 
\end{tabular}{\scriptsize\par}
\end{table}

\section{Conclusions:}

In conclusion, we have reported the ground- and excited-state properties
of the RP phases of A$_{n+1}$Hf$_{n}$S$_{3n+1}$ (where A = Ca,
Sr, Ba and $n$ = 1$-$3) using a combination of density functional
theory and many-body perturbation theory. Our study demonstrates that
these RP phases are mechanically stable and exhibit ductile behavior.
The G$_{0}$W$_{0}$@PBE bandgaps (1.43$-$2.14 eV) of these RP phases
are smaller than those of their bulk counterparts, while the low effective
masses of electrons suggest excellent charge carrier mobility, beneficial
for energy-harvesting applications. BSE and mBSE calculations further
reveal that these materials have an optical bandgap in the visible
region, with exciton binding energies (0.17$-$0.33 eV) decreasing
as the perovskite layer thickness increases. We also observe a dominant
electronic contribution and a negligible ionic contribution to the
effective dielectric screening in these RP phases. In addition, by
calculating the electron-phonon coupling parameters, we find that
these compounds exhibit significant carrier-lattice interactions,
with charge-separated polaronic states being less stable than the
bound excitons. The Hellwarth polaron model predicts that these RP
phases exhibit higher polaronic mobility of electrons (26.32$-$121.32
cm$^{2}$V$^{-1}$s$^{-1}$) compared to their bulk counterparts.
Overall, these results are anticipated to accelerate the research
and adoption of RP phases in optoelectronic applications.
\begin{acknowledgments}
S.A. would like to acknowledge the Council of Scientific and Industrial
Research (CSIR), Government of India {[}Grant No. 09/1128(11453)/2021-EMR-I{]}
for Senior Research Fellowship. The authors acknowledge the High Performance
Computing Cluster (HPCC) \textquoteleft Magus\textquoteright{} at
Shiv Nadar Institution of Eminence for providing computational resources
that have contributed to the research results reported within this
paper.
\end{acknowledgments}

\bibliographystyle{apsrev4-2}
\bibliography{refs}

\end{document}